\documentclass[preprint,showpacs,aps]{revtex4}

\usepackage{amsmath}
\usepackage{dcolumn}
\usepackage{verbatim}
\usepackage{mathrsfs}

\begin{document}

\title{Shell-model test of the rotational-model relation between static
quadrupole moments $Q(2^+_1)$, $B(E2)$'s, and orbital $M1$ transitions}

\author{S.~J.~Q.~Robinson}
\affiliation{Geology and Physics Department, University of Southern 
Indiana, Evansville, Indiana 47712}

\author{L.~Zamick and A. Escuderos}
\affiliation{Department of Physics and Astronomy, Rutgers University,
Piscataway, New Jersey 08854}
 
\author{R.~W.~Fearick}
\affiliation{Physics Department, University of Cape Town, Rondebosch 7700,
South Africa}

\author{P.~von~Neumann-Cosel and A.~Richter}
\affiliation{Institut f\"ur Kernphysik, Technische Universit\"at Darmstadt,
D-64289 Darmstadt, Germany}

\date{\today}

\begin{abstract} 

In this work, we examine critically the relation between orbital magnetic 
dipole (scissors mode) strength and quadrupole deformation properties. Assuming
a simple $K=0$ ground state band in an even--even nucleus, the quantities
$Q(2^+_1)$ (i.e., the static quadrupole moment) and $B(E2)_{0_1 \to 2_1}$ both
are described by a single parameter---the intrinsic quadrupole  moment $Q_0$.
In the shell model, we can operationally define $Q_0 (\text{Static})$ and
$Q_0(BE2)$ and see if they are the same. Following a brief excursion to the
$sd$ shell, we perform calculations in the $fp$ shell. The nuclei we consider
($^{44,46,48}$Ti and $^{48,50}$Cr) are far from being perfect  rotors, but we
find that the calculated ratio $Q_0(\text{Static})/Q_0(BE2)$ is in many cases
surprisingly close to one. We also discuss the  quadrupole collectivity of 
orbital magnetic dipole transitions. We find that  the large orbital $B(M1)$ 
strength in $^{44}$Ti relative to $^{46}$Ti and  $^{48}$Ti cannot be explained 
by simple deformation arguments. 

\end{abstract}


\maketitle

\section{Introduction}

In this work we will make a comparison of the shell model and the collective 
model for several quantities that are sensitive to nuclear deformation. These 
include $B(E2)$'s, static quadrupole moments, and orbital magnetic dipole 
transitions. This will be a theory versus theory work. Some experimental 
results are quoted and serve as anchors for our results, but we will not be 
inhibited by the lack of experimental data in doing these calculations. We plan
in the near future to make a more extensive theory--experiment comparison. But 
there are holes in the experimental data which must be filled.

The main thrust of our work will be to understand the relationship of orbital
magnetic dipole transitions to quadrupole deformations in the nucleus. For 
example, after the experimental discovery in heavy deformed nuclei of the
relation between the orbital magnetic dipole strength and nuclear
deformation~\cite{zrrs90}, there have been many works which relate the orbital
$M1$ (scissors mode) strength to electric quadrupole transition rates
($B(E2)_{0_1 \to 2_1}$), often assuming that they are proportional to each
other~\cite{hcrw92,zz92,ir93,ngbr95,eknrr99}. 

First, though, we shall do survey calculations of $B(E2)$'s and static 
quadrupole moments in the $fp$ shell to see how well the shell model relates to
the simple rotational model of Bohr and Mottelson~\cite{bm75}. In this 
rotational model, the formulae for $B(E2)$'s and static quadrupole moments 
involve a single parameter---the intrinsic quadrupole moment. These formulae 
are, respectively,
\begin{subequations}
\begin{equation}
B(E2) = \frac{5}{16 \pi} Q^2_0(B) \left| \langle I_1 K 2 0 | I_2 K \rangle
\right|^2 
\end{equation}
\begin{equation}
Q(I) = \frac{3 K^2 - I(I+1)}{(I+1)(2I+3)} Q_0(S) ,
\end{equation}
\end{subequations}
where $B$ and $S$ stand for $B(E2)$ and ``Static'', respectively. Here $Q_0$ is
the intrinsic quadrupole moment---what we would see in the rotational frame. On
the other hand, $Q(I)$ is what we measure in the laboratory. In the simple
rotational model, $Q_0(B)$ is equal to $Q_0(S)$.

For the case $I_1=0$, $I_2=2$, the Clebsch-Gordan coefficient above is 1. For a
simple $K=0$ band in an even--even nucleus, we obtain
\begin{subequations}
\begin{equation}
B(E2)_{0 \to 2} = \frac{5}{16 \pi} Q^2_0 (B) 
\end{equation}
\begin{equation}
Q(2^+) = -\frac{2}{7} Q_0 (S) .
\end{equation}
\end{subequations}
Note that the laboratory quadrupole moment has the opposite sign of the 
intrinsic quadrupole moment---a well known result. It can be understood 
physically by imagining rotating a cigar (which has a positive quadrupole 
moment, i.e., prolate) about an axis perpendicular to the line of the cigar.
This will trace out a flat pancake shape which is oblate.

We then find that the ratio
\begin{equation}
\frac{Q_0(S)}{Q_0(B)} = -\frac{7}{2} \sqrt{\frac{5}{16 \pi}} \frac{Q(2^+)}
{\sqrt{B(E2)}} = -1.1038705 \frac{Q(2^+)}{\sqrt{B(E2)}} . \label{ratio}
\end{equation}

\section{Quadrupole properties in the $sd$ shell}

Although we will be performing calculations in the $fp$ shell, we shall here
briefly look over the experimental situation in the $sd$ shell. In
Table~\ref{tab:sdshell} we show experimental values of $Q(2^+_1)$~\cite{s01},
$B(E2)$~\cite{rnt01}, and the ratio $|Q_0(S)/Q_0(B)|$ as given by
Eq.~\eqref{ratio}. We also show the experimental values of $E(4^+_1)/E(2^+_1)$
as a measure of how close we are to the rotational limit of $10/3$, or the
vibrational limit of $2/1$.

\begin{table}[ht] 

\caption{Experimental data on $Q(2^+_1)$ and $B(E2)$ in the $sd$ shell and the
ratio $|Q_0(S)/Q_0(B)|$; we also give the experimental values of $E(4)/E(2)$.}
\label{tab:sdshell} 

\begin{tabular*}{.75\textwidth}[t]{@{\extracolsep{\fill}}ldddd}
\toprule
 & \multicolumn{1}{c}{$Q(2^+_1)$} & \multicolumn{1}{c}{$B(E2)$} &
 \multicolumn{1}{c}{$|Q_0(S)/Q_0(B)|$} & 
 \multicolumn{1}{c}{$E(4^+_1)/E(2^+_1)$} \\ 
 & \multicolumn{1}{c}{[e fm$^2$]} & \multicolumn{1}{c}{[e$^2$ fm$^4$]} \\ 
\colrule
$^{20}$Ne & -23 & 340 & 1.377 & 2.600 \\
$^{22}$Ne & -19 & 230 & 1.383 & 2.634 \\
$^{24}$Mg & -16.6 & 432 & 0.802 & 3.012 \\
$^{28}$Si & +16.5 & 320 & 1.018 & 2.596 \\
$^{32}$S & -14.9 & 300 & 0.950 & 2.000 \\
$^{36}$Ar & +11 & 340 & 0.658 & 2.240 \\
$^{40}$Ar$^*$ & +1 & 330 & 0.061 & 1.980 \\
\botrule
\end{tabular*}

\end{table}

Note that the static quadrupole moments of $^{20}$Ne, $^{22}$Ne, $^{24}$Mg, and
$^{32}$S are negative, while those of $^{28}$Si and $^{36}$Ar are positive. If
we limit ourselves to axial symmetry, this indicates that the first group has
prolate ground state bands and the second group has oblate ones. Skyrme II
Hartree-Fock results by Jaqaman and Zamick~\cite{jz84} correctly give the signs
of all the static quadrupole moments. The small static quadrupole moment of
$^{40}$Ar is consistent with magnetic moment results of the $2^+_1$ state by
Stefanova et al.~\cite{setal05}.

The ratio $|Q_0(S)/Q_0(B)|$ for $^{20}$Ne is {\it larger} than the rotational
limit, 1.377 versus 1; likewise $^{22}$Ne. In the case of $^{24}$Mg,
$|Q_0(S)/Q_0(B)|$ is smaller than for $^{20}$Ne or $^{22}$Ne, despite the fact
that the spectrum is closer to rotational for $^{24}$Mg. Also surprisingly for
$^{32}$S, the ratio $E(4)/E(2)$ is 2.000, the vibrational limit, for which one
might expect a near zero static quadrupole moment. But the ratio
$|Q_0(S)/Q_0(B)|$ is 0.950, close to the simple rotational prediction of unity.

In general, it is difficult to correlate $|Q_0(S)/Q_0(B)|$ with $E(4)/E(2)$
assuming a simple axially symmetric rotor.

We should mention that an analysis of the relationship of $Q_0(S)$ and $Q_0(B)$
has already been performed by Bender, Flocard, and Heenen~\cite{bfh03} and
Bender et al.~\cite{bbdh04}, albeit not for the $fp$-shell nuclei considered 
here and using a different method. They perform angular momentum projections on
BCS--Hartree-Fock states obtained with the Skyrme interaction SLy6 for the
particle--hole channel and a density-dependent contact force in the pairing
channel~\cite{bfh03}. Their calculations are mainly in the $sd$ 
shell~\cite{bfh03} and neutron-deficient lead region~\cite{bbdh04}. For one
nucleus in common, $^{40}$Ca, their results for $0p-0h$, $2p-2h$, $4p-4h$,
$6p-6h$, $8p-8h$, and $12p-12h$ do not differ so much from previous 
calculations of Zheng, Berdichevsky, and Zamick~\cite{zbz88} as far as the
intrinsic properties are concerned, but their calculation has the added feature
of providing an energy spectrum and expectation values in the laboratory frame.

In Ref.~\cite{jz84}, the authors predict that $^{36}$Ar is oblate. This is
confirmed by the fact that the static quadrupole moment of the $2^+_1$ state is
positive: $+11$~e fm$^2$~\cite{s01}. The experimental $B(E2)$ is 340~e$^2$
fm$^4$ and $|\beta_2|=0.273$~\cite{rnt01}. Using Eq.~\eqref{ratio}, we find
\begin{equation}
\frac{Q_0(S)}{Q_0(B)} = 0.6505236 ~.
\end{equation}
The energy ratio is
\begin{equation}
\frac{E(4^+_1)}{E(2^+_1)} = \frac{4414.36}{1970.35} = 2.240 ~.
\end{equation}
These results are consistent with a nucleus not being too rotational.

The corresponding numbers in the calculation of Bender et al.~\cite{bfh03} are
\begin{equation}
Q(2^+_1)_{\text{lab}} = 13 ~\text{e fm$^2$} ~, \hspace{1.5cm}
B(E2)\uparrow = 220 ~\text{e$^2$ fm$^4$} ~, \hspace{1.5cm}
\beta = -0.21 ~.
\end{equation}
The calculated ratios are
\begin{equation}
\frac{Q_0(S)}{Q_0(B)} = 0.9675 ~, \hspace{2cm}
\frac{E(4^+_1)}{E(2^+_1)} = 2.6545 ~.
\end{equation}
These calculations~\cite{bfh03} give a more rotational picture than experiment.
There is a consistency, however, in that a larger ratio $E(4)/E(2)$ yields a
larger ratio $Q_0(S)/Q_0(B)$.

\section{Shell model calculations of $B(E2)$ and $Q(2^+_1)$, and how they
relate to the simple rotational model}
%{A shell model test}

We will put the above relation~(\ref{ratio}) to the test in a shell model 
approach for the following nuclei: $^{44}$Ti, $^{46}$Ti, $^{48}$Ti, $^{48}$Cr, 
and $^{50}$Cr. We use the OXBASH program~\cite{eetal85} and the FPD6 
interaction~\cite{rmjb91}.

The nuclei that we have chosen are far from being perfect rotors. Their
description falls somewhere between vibrational and rotational. The ratios 
$E(4)/E(2)$, which would all be $10/3$ in the simple rotational case, are as
follows: 1.922, 2.010, 2.118, 2.459, 2.342, for $^{44}$Ti, $^{46}$Ti, 
$^{48}$Ti, $^{48}$Cr, and $^{50}$Cr, respectively.

We perform shell model calculations in a complete $fp$ space using the FPD6
interaction. We assign effective charges of 1.5 for the protons and 0.5 for
the neutrons. We calculate $B(E2)_{0_1 \to 2_1}$ and $Q(2^+)$ (the laboratory
$Q$, of course) and put them into Eq.~(\ref{ratio}) in order to get operational
values of $Q_0(S)/Q_0(B)$. The results are given in Table~\ref{tab:ratios}.

\begin{table}[ht]
\caption{The quantity $Q_0(S)/Q_0(B)$ as obtained in the shell model with the
FPD6 interaction and experiment.} \label{tab:ratios}
\begin{tabular*}{.7\textwidth}{@{\extracolsep{\fill}}cddd}
\toprule
 & \multicolumn{1}{c}{$Q(2^+_1)$} & \multicolumn{1}{c}{$B(E2)$} & 
\multicolumn{1}{c}{$Q_0(S)/Q_0(B)$} \\
 & \multicolumn{1}{c}{[e fm$^2$]} & \multicolumn{1}{c}{[e$^2$ fm$^4$]} \\ 
\colrule
Theory (FPD6): \\ \cline{1-1}
$^{44}$Ti & -20.156 & 607.24 & 0.9029 \\
$^{46}$Ti & -22.071 & 682.06 & 0.9329 \\
$^{48}$Ti & -17.714 & 560.78 & 0.8257 \\
$^{48}$Cr & -33.271 & 1378.4 & 0.9892 \\
$^{50}$Cr & -30.955 & 1219.0 & 0.9787 \\
Experiment: \\ \cline{1-1}
$^{44}$Ti & & 650 & \\
$^{46}$Ti & -21 & 950 & 0.7521 \\
$^{48}$Ti & -17.7 & 720 & 0.7282 \\
$^{48}$Cr & & 1360 & \\
$^{50}$Cr & -36 & 1080 & 1.2092 \\
\botrule
\end{tabular*}
\end{table}

Except for $^{48}$Ti, the FPD6 results for the ratios are all greater than 
$0.9$, reading a maximum of $0.9892$ for $^{48}$Cr. It is somewhat surprising 
that these ratios are so close to 1, given that the ratios $E(4)/E(2)$ are much
further away from the rotational limit $10/3$.

We can also obtain some of the above ratios from experiment. We refer to the 
compilation of nuclear moments of Stone~\cite{s01} and of $B(E2)$'s by Raman
et~al.~\cite{rnt01}. Taking these experiments at face value, we see that the 
ratio $Q_0(S)/Q_0(B)$ reduces to about 0.75 for $^{46}$Ti and $^{48}$Ti, but
is bigger than 1 for $^{50}$Cr.

It should be noted that in the simplest version of the vibrational mode, 
$Q_0(S)$ is zero. We can imagine a nucleus vibrating between a prolate shape 
and an oblate shape, and causing the quadrupole moment to average to zero. On 
the other hand, the $B(E2)_{0_1 \to 2_1}$ is quite large in this vibrational 
limit, causing the ratio $Q_0(S)/Q_0(B)$ to be zero or, in more sophisticated 
vibrational models, quite small.

\section{Results with an alternate interaction T0FPD6}

For systems of identical particles, e.g., the tin isotopes, which in the simple
shell model involve only valence neutrons, one does not get rotational 
behaviour. One does go closer to the rotational limit when one has many 
open-shell neutrons and protons. Whereas two identical nucleons must have 
isospin 1, a neutron and a proton can have both isospin 0 and 1. This suggests 
that the $T=0$ part of the nucleon--nucleon interaction plays an important role
in enhancing nuclear rotational collectivity.

In this section, we will use an interaction TOFPD6 that is the same as the FPD6
interaction for $T=1$ states, but vanishes for $T=0$ states. We thus expect 
that the rotational collectivity will be reduced. This interaction serves as a 
counterpoint of the full interaction in the previous section. It has been
discussed before by Robinson and Zamick~\cite{rz01}.

We present the results for T0FPD6 in Table~\ref{tab:t0fpd6}.

\begin{table}[ht]
\caption{The same as Table~\ref{tab:ratios}, but with the $T=0$ two-body matrix
elements of FPD6 set to zero (T0FPD6).} \label{tab:t0fpd6}
\begin{tabular*}{.6\textwidth}{@{\extracolsep{\fill}}cddd}
\toprule
Nucleus & \multicolumn{1}{c}{$Q(2^+_1)$} & \multicolumn{1}{c}{$B(E2)$} & 
\multicolumn{1}{c}{$Q_0(S)/Q_0(B)$} \\
 & \multicolumn{1}{c}{[e fm$^2$]} & \multicolumn{1}{c}{[e$^2$ fm$^4$]} \\ 
\colrule
$^{44}$Ti & -0.880 & 375.09 & 0.0502 \\
$^{46}$Ti & -8.195 & 432.81 & 0.4348 \\
$^{48}$Ti & -8.777 & 401.97 & 0.4832 \\
$^{48}$Cr & -21.437 & 813.06 & 0.8299 \\
$^{50}$Cr & -20.985 & 736.60 & 0.8535 \\
\botrule
\end{tabular*}
\end{table}

We see that both $Q(2_1^+)$ and $B(E2)$ decrease in magnitude, consistent with 
the above discussion. However, $Q_0(S)$ decreases more rapidly than $\sqrt{
B(E2)}$, so the ratio $Q_0(S)/Q_0(B)$ is less for this case than when the full 
interaction is present. For $^{44}$Ti this ratio is 0.0502, close to the 
vibrational limit of zero. For $^{48}$Cr the ratio decreases from 0.9892 to 
0.8299; nevertheless, it is still substantial, indicating that the $T=1$ 
interaction, acting alone, can lead us to some extent in the direction of the 
rotational limit.

\section{Random interaction studies}

Nuclei in the region we are considering have undergone Random Interaction
studies. We refer to the works of Vel\'azquez et al.~\cite{vhfz03} and 
Zelevinsky and Volya~\cite{zv04}. These works were stimulated by that of 
Johnson, Bertsch, and Dean~\cite{jbd98}. 

In particular, in the work of Zelevinsky, a quantity is considered which is 
proportional to the square of $Q_0(S)/Q_0(B)$ and which is normalized to 
$(2/7)^2$ if $Q_0(S)/Q_0(B)$ equals 1. He calls this the {\it Alaga ratio}. He 
selects cases for which the random interaction yields a $J=0,~J=2$~ sequence of
lowest energies. For these he finds two peaks, one corresponding to $Q_0(S)/
Q_0(B)=1$ and the other to $Q_0(S)/Q_0(B)=0$. In our terminology, these would
correspond to the rotational limit in the former case and either the simple
vibrational limit or the spherical limit in the latter case. He cites early
work of R.~Rockmore~\cite{r61} as affording an explanation of this surprising
behaviour.

\section{Orbital magnetic dipole transitions in $^{44}$T\lowercase{i},
$^{46}$T\lowercase{i}, and  $^{48}$T\lowercase{i}}

The orbital magnetic isovector dipole transitions, i.e., scissors mode 
excitations, also display collective behaviour~\cite{zrrs90}. There are 
systematics which suggest that $B(M1)_{\text{orbital}}$ is roughly proportional
to $B(E2)$. There are more detailed sophisticated relations as well. If one 
uses a simple quadrupole--quadrupole interaction, the energy
weighted $B(M1)_{\text{orbital}}$ is proportional to the difference $\Big(
B(E2)_{\text{isoscalar}}-B(E2)_{\text{isovector}}\Big)$~\cite{zz92}.

The bare orbital $M1$ operator is
\begin{equation}
\sqrt{\frac{3}{4 \pi}} \sum l(i) g_l(i) ,
\end{equation}
where $g_l$ is 1 for a proton and 0 for a neutron. This is the operator that
we use in the calculations.

How to extract the scissors mode strength is not completely unambiguous. The 
mode is associated with low-lying $1^+$ excitations at around 3~MeV. But the 
strength can be fragmented even at this lowest energy. Besides this, there is 
orbital strength at higher energies, a somewhat grassy behaviour where 
individual states are very weakly excited but, because there are so many of 
them, the total orbital strength can be significant.

Therefore, we will give three sets of values (see Table~\ref{tab:bm1ti}). 
First, we give the strength to  the lowest state, then to the lowest 10 states,
and finally to the lowest 1000 states (except for $^{48}$Ca, where we include
only 300 states). The 10-states strength should encompass what we usually call
the  scissors mode, while the 1000-states strength is close to the total
strength  including the grassy, non-collective part. It would appear that the
highest excitation energies reached in the experiments~\cite{zrrs90,gdrjvw90}
are not sufficient to reach the $T+1$ part of the spectrum.

\begin{table}[ht]
\caption{The calculated orbital $M1$ strengths ($\mu_\text{N}^2$). Unless 
indicated, the calculations are made with the FPD6 interaction.} 
\label{tab:bm1ti}
\begin{tabular*}{.9\textwidth}{@{\extracolsep{\fill}}lddddd}
\toprule
 & \multicolumn{1}{c}{$^{44}$Ti} & \multicolumn{1}{c}{$^{46}$Ti} & 
\multicolumn{1}{c}{$^{48}$Ti} & \multicolumn{1}{c}{$^{48}$Ti$/^{46}$Ti} &
\multicolumn{1}{c}{$^{48}$Cr} \\
\colrule
$T \to T$ \\ \cline{1-1}
Lowest state & 0.0017 & 0.305 & 0.105 & 0.3443 \\
Lowest 10 states & 0.0320 & 0.5979 & 0.3056 & 0.5111 \\
Lowest 100 states & & 0.79 & 0.504 & 0.6380 \\ \cline{2-5}
All states & 0.0355 & 0.9195 & 0.7191 & 0.7820 \\ \cline{2-5}
All states (T0FPD6) & 0.0583 & 0.2166 & 0.2235 & 1.0319 \\ \\
$T \to T+1$ \\ \cline{1-1}
Lowest state & 0.862 & 0.0991 & 0.0041 & 0.4450 & 0.784 \\
Lowest 10 states & 1.4317 & 0.368 & 0.1951 & 0.5302 & 1.3855 \\
Lowest 100 states & 2.12 & & & & 1.994 \\ \cline{2-6}
All states & 2.127 & 0.5616 & 0.3099 & 0.5518 & 2.271\footnote{Lowest 300
states} \\ \cline{2-6}
All states (T0FPD6) & 1.1118 & 0.3940 & 0.1671 & 0.4241 \\
\botrule
\end{tabular*}
\end{table}

We first discuss the nuclei $^{46}$Ti and $^{48}$Ti, for which there are some 
data on $B(M1)$. We see consistently that the orbital $B(M1)$ strength is 
larger in $^{46}$Ti than in $^{48}$Ti. This is consistent with the fact that 
$^{46}$Ti has a greater $B(E2)$ and static $2^+$ quadrupole moment than 
$^{48}$Ti.

We next consider the $N=Z$ nucleus $^{44}$Ti, for which there is no data 
because this nucleus is unstable. The isoscalar orbital $B(M1)$ strength is 
very weak. This is also true for the spin $B(M1)$, but for a different reason. 
The isoscalar spin coupling is much smaller than the isovector one. For the 
orbital case, the couplings are equal because the operator is $\sum_{\text{
protons}} \vec{\ell}$. For the orbital case, the $B(M1)$ isoscalar is very weak
because the correlations due to the nuclear interaction move the ground state 
towards the $SU(4)$ limit, in which $LS$ coupling holds and for which the 
ground state is a pure $L=0$ state. For this extreme case, the $B(M1)$ orbital 
isoscalar will vanish.

The transitions of interest for $^{44}$Ti are, therefore, the isovector orbital
dipole ones. The ($T \to T+1$) $B(M1)_{\text{orbital}}$ summed strength is 
larger in $^{44}$Ti than the ($T \to T $) and [$(T \to T) ~+~ (T \to T+1)$]
strengths in $^{46}$Ti and $^{48}$Ti, which are 0.5615 (1.4810) and 0.3099 
(1.0288) $\mu_{\text{N}}^2$, respectively.  On the other hand, $^{44}$Ti is not
more deformed than $^{46}$Ti. According to Raman et al.~\cite{rnt01}, the
values of the quadrupole deformation parameters $\beta$ for $^{44,46,48}$Ti and
$^{48,50}$Cr are, respectively, 0.27, 0.317, 0.269, 0.335, and 0.293. Thus, we
have here in $fp$-shell nuclei a counter-example to the experimentally
established proportionality between the orbital $B(M1)$ and the $B(E2)$ in
heavy deformed nuclei. Perhaps there are correlations which cause an
enhancement for $N=2$ nuclei.

An analysis by Retamosa et al.~\cite{rupm90} in the $sd$ shell comparing
$^{20}$Ne, $^{22}$Ne, and $^{24}$Mg was performed (somewhat analogous to
$^{44}$Ti and $^{46}$Ti for the first two cases), but no anomaly was reported
there. In the $SU(3)$ model, they found consistency in the relation of $B(M1)
_{\text{orbital}}$ to deformation. In this limit, the summed $M1$ strengths 
(all orbital) for $^{20}$Ne, $^{22}$Ne, and $^{24}$Mg were 1.1, 1.17, and 1.6 
$\mu_\text{N}^2$, respectively. Looking at the Raman tables~\cite{rnt01} for
these nuclei, there is some complication---the deformation parameters $\beta_2$
are not in one-to-one correspondence with the $B(E2)$'s. The values of
$(B(E2), \beta)$ for these three nuclei from the Raman tables~\cite{rnt01} are,
respectively, $(0.034,0.728)$, $(0.0236, 0.562)$, and $(0.0432, 0.606)$, where
the units for $B(E2)$ are b$^2$. The authors also do calculations with a more 
realistic interaction, but no enough strengths are listed in order to make a 
comparison for the point we are trying to make. Retamosa et al.~\cite{rupm90}
also give strengths to the first $10^+$ states in $^{44}$Ti; our numbers are
consistent with theirs.

Earlier works on the shell model for light nuclei include L.~Zamick~\cite{z85}
and A.~Poves~\cite{prm89}.

\section{Closing remarks}

In this work we have examined what predictions the shell model make for
collective properties which are after dealt with in the rotational model. 
Although the nuclei are far from perfect rotors, the calculated ratio 
$Q_0(S)/Q_0(B)$ is fairly close to 1 in many cases. When the FPD6 interaction
is used, the orbital magnetic dipole transitions for $^{46,48}$Ti also fit into
this picture, although there is the added complication of separating the 
collective from the non-collective part in this case. Also there is a 
substantial enhancement for the $N=Z$ nucleus $^{44}$Ti, which cannot be 
explained as purely a deformation effect.  We hope our work will stimulate more
experimental investigations. There is information of $B(M1)$ rates in $^{46}$Ti
and $^{48}$Ti, but thus far the orbital $B(M1)$ has  only been extracted in
$^{48}$Ti. However, in a short time, we will be able to  make a more extensive
theory--experiment study of these magnetic dipole  transitions.

\begin{acknowledgments}
This work has been supported by the DFG under contracts SFB 634 and 445 
SUA-113/6/0-1, and by the NRF, South Africa. AE acknowledges support from the 
Secretar\'{\i}a de Estado de Educaci\'on y Universidades (Spain) and the 
European Social Fund. We thank Y.~Y.~Sharon for his help.
\end{acknowledgments}

\end{document}